\documentstyle[12pt,epsf]{article}
\newcommand{\eq}{\begin{equation}}
\newcommand{\en}{\end{equation}}
\newcommand{\eqa}{\begin{eqnarray}}
\newcommand{\ena}{\end{eqnarray}}
\newcommand{\half}{\frac{1}{2}}
\newcommand{\ra}{\rightarrow}

\newcommand{\tr}{\mbox{Tr}}
\newcommand{\noi}{\noindent}

\newcommand{\bea}{\begin{eqnarray}}
\newcommand{\eea}{\end{eqnarray}}
\newcommand{\AmS}{{\protect\the\textfont2
  A\kern-.1667em\lower.5ex\hbox{M}\kern-.125emS}}
\begin{document}
\noindent
\begin{titlepage}
\begin{flushright}
SHEP 95/42\\
KANAZAWA 96-02\\
HUB-EP-96/6\\
HLRZ 77/95\\
WUB 95-45\\
\end{flushright}
\begin{center}
{\Large\bf Dual Superconductor Scenario of Confinement:\\[2mm]
A Systematic Study of Gribov Copy Effects}\\[1cm]

{\bf G.S.\ Bali}\\
{\small\it Physics Department, The University, Highfield,
Southampton SO17 1BJ, UK}\\[3mm]
{\bf V.\ Bornyakov}\\
{\small\it Department of Physics,  Kanazawa University, Kanazawa
920-11, Japan \\[1mm]
{\rm and} Institute for High Energy Physics,
142284 Protvino, Russia}\footnote{Permanent address.}  \\[3mm]
{\bf M.\ M\"uller-Preussker}\\
{\small\it Institut f\"ur Physik,
Humboldt Universit\"at zu Berlin, 10099 Berlin, Germany}\\[3mm]
{\bf K.\ Schilling}\\
{\small\it Fachbereich Physik, Bergische Universit\"at,
Gesamthochschule, 42097 Wuppertal, Germany \\[1mm]
{\rm and} HLRZ c/o KFA, 52425 J\"ulich, {\rm and}
DESY, 22603 Hamburg, Germany}\\[1cm]
\end{center}

\begin{abstract}
\noindent We perform a study of the effects from maximal
abelian gauge Gribov copies in the context of the dual superconductor
scenario of confinement, on the basis of a novel approach for estimation
of systematic uncertainties from incomplete gauge fixing. We present
numerical results, in $SU(2)$ lattice gauge theory, using the
overrelaxed simulated annealing gauge fixing algorithm. We find
abelian and non-abelian string tensions to differ significantly, their
ratio being $0.92(4)$ at $\beta = 2.5115$.  An approximate
factorization of the abelian potential into monopole and photon
contributions has been confirmed, the former giving rise to the
abelian string tension.
\end{abstract}

\end{titlepage}
\section{Introduction}

Understanding how confinement arises from QCD is a central problem of
strong interaction physics. G.~'t~Hooft and S.~Mandelstam proposed
the QCD vacuum state to behave like a magnetic
superconductor~\cite{hooft1,mand1}.  A dual Meissner effect is
believed to be responsible for the formation of thin string-like
chromo-electric flux tubes between quarks in $SU(N)$ Yang-Mills
theories.  This confinement mechanism has indeed been established in
compact QED~\cite{poly,banks,smitvds}.  The disorder of the related
topological objects --- magnetic monopoles --- gives rise to an area
law for large Wilson loops and, thus, leads to a confining potential.
The application of this idea to non-abelian gauge theories is based on
the abelian projection~\cite{hooft2}, reducing the non-abelian
$SU(N)$ gauge symmetry to the maximal abelian (Cartan) subgroup
$U(1)^{N-1}$ by gauge fixing the off-diagonal components of the gauge
field.  Then the theory can be regarded as an abelian gauge theory
with magnetic monopoles
and charged matter fields (quarks and off-diagonal gluons).
The dual superconductor idea is realized if
these abelian monopoles condense. In this scenario large distance
(low momentum)
properties of QCD are carried by the abelian degrees of freedom
(abelian dominance).

Nonperturbative investigations of this conjecture became possible after
formulating the abelian projection for lattice gauge theories~\cite{desy1}.
In previous numerical studies (see
reviews~\cite{suzrev,dgrev} and references therein), it has been demonstrated
that the maximal abelian (MA) gauge was
a very suitable
candidate for
lattice investigations of the 't~Hooft-Mandelstam confinement scenario,
being the only known renormalizable abelian gauge.

These studies
provided strong evidence for
the QCD vacuum indeed to act like a dual superconductor.
In Ref.~\cite{suz1} the
value of the abelian string tension, i.e.\ the slope of
the
static potential between charge one static sources
at large distance,
has been observed
to be close to that of
the potential between
static quarks in the fundamental representation  of the non-abelian
theory.
This feature supports abelian dominance as predicted by 't Hooft.
Recently, various results in favor
of this picture have been obtained by other groups as well~\cite{dgrev}.

At this stage it appears to be important to address such issues on a
more quantitative level. This requires a careful study of the problem of
gauge (or Gribov) ambiguities~\cite{grib},
and the resulting biases on abelian observables.
In the present paper we
aim at removing this uncertainty of previous studies.
We shall develop a
new effective gauge fixing algorithm, thus reducing gauge
fixing ambiguities with respect to the standard relaxation algorithm
employed so far. We propose a numerical procedure to estimate the
remaining gauge fixing biases. This enables us to carry out high
precision measurements of the abelian string tension and other abelian
observables, with control over systematic errors. The main results
have been obtained on a $32^4$ lattice at $\beta = 2.5115$.

Our investigation revealed that, compared to typical statistical
errors, the effect of gauge copies cannot be neglected, even with the
improved gauge fixing algorithm.
Nevertheless, the new algorithm does reduce the variance of
observables with respect to various gauge copies
considerably, compared
with the traditional  relaxation plus overrelaxation algorithm, and, in
average, yields larger values of the functional to be maximized.

Our paper is organized as follows.  In Section~2 we shortly review the
status of gauge fixing and monopole kinematics to set the stage and to
present the underlying formulae in a self contained fashion. In
Section~3, we present the overrelaxed simulated annealing gauge fixing
algorithm and compare it to the standard algorithm. In Section~4 we
elaborate on  a procedure to estimate the effect of remaining
gauge fixing ambiguities on abelian observables, such as the monopole
density or the abelian potential.  Physical results on the non-abelian
and abelian potentials, the decomposition of the potential
into monopole and photon contributions, the abelian potential
between charge two static sources and the monopole density
are presented in Section~5.

\section{Physics from maximal abelian gauge}

\subsection{Abelian projection}
We start from the lattice version~\cite{desy1} of 't Hooft's abelian
projection~\cite{hooft2}.  The idea is to partially fix gauge degrees
of freedom such that the maximal abelian (Cartan) subgroup
($U^{N-1}(1)$ for $SU(N)$ gauge groups) remains unbroken.

A few abelian gauges have been suggested in Ref.~\cite{hooft2} where
MA gauge, referring to a differential gauge condition, has been
favored.  Lattice simulations have indeed demonstrated MA gauge to be
very suitable for investigations of abelian projections of gauge
theories.

In the following, we will restrict ourselves to the case of $SU(2)$
gauge theory.
Fixing MA gauge on the lattice amounts to maximizing
the functional ($V=N_{\mbox{\scriptsize sites}}$),
\eq
 F(U) = \frac{1}{8 V} \sum_{n,\mu}
        \tr\left( \sigma_3 U_{n,\mu} \sigma_3
      U_{n,\mu}^{\dagger} \right)\quad,                \label{maxfunc}
\en
\noindent
with respect to local gauge transformations,

\eq
U_{n,\mu} \ra U^{g} _{n,\mu} = g_n U_{n,\mu} g_{n+
\hat{\mu}}^{\dagger}\quad.    \label{201}
\en

\noi
Condition Eq.~(\ref{maxfunc}) fixes (besides other possible degeneracies)
$g_n$ only up to multiplications
$g_n \rightarrow v_n  g_n$
with $v_n = \exp(i\alpha_n\tau_3)$, $\tau_3=\sigma_3/2$, $-2 \pi \leq
\alpha_n < 2 \pi$, i.e.\ $g_n\in SU(2)/U(1)$.

It has been shown that the corresponding continuum gauge, defined
by the bilinear differential gauge condition,
\eq
(\partial_{\mu} \pm igA^{3}_{\mu})A^{\pm}_{\mu}=0\quad,\quad
A^{\pm}_{\mu}=A^{1}_{\mu} \pm iA^{2}_{\mu}\quad,    \label{cgc}
\en

\noi is renormalizable~\cite{renorm}, a feature, that is crucial
 for a continuum interpretation of lattice results.

After a configuration has been transformed to satisfy the
MA gauge condition, a coset decomposition is
performed,
\eq
U_{n,\mu} = C_{n,\mu} V_{n,\mu}\quad ,
\label{coset}
\en

\noi
where $V_{n,\mu} = \exp(i \phi_{n,\mu} \tau_3),~
-2 \pi \leq \phi_{n,\mu} < 2 \pi$,
transforms like a (neutral) gauge field and $C_{n,\mu}$ like
a charged matter field
with respect to transformations within the residual abelian subgroup,
\eq
V_{n,\mu} \ra v_n V_{n,\mu}v_{n+\hat{\mu}}^{\dagger}\quad,\quad
C_{n,\mu}  \ra v_n C_{n,\mu} v_n ^{\dagger}\quad.
\label{abgtr}
\en

\noi
Quark fields are also charged with respect to such $U(1)$ transformations.
The abelian lattice gauge fields $V_{n,\mu}$ constitute an abelian
projected configuration.

The $SU(2)$ action of the original gauge theory can be decomposed into
a $U(1)$ pure gauge action, a term describing interactions of the $U(1)$ gauge
fields with charged fields, i.e. the off-diagonal components,
and a self-interaction term of those charged fields~\cite{polik_vap}.
Maximizing the diagonal components of all gauge fields with respect to
the off-diagonal components amounts to enhancing the
effect of the pure $U(1)$ gauge part in comparison with those contributions
containing interactions with charged fields. On a heuristic level,
one might expect the importance of the
$U(1)$ degrees of freedom to be enhanced in the
MA projected theory, in comparison with other abelian projections.

The abelian Wilson loop for charge one static sources is defined as

\eq
W^{ab}(C) = \half \tr\left(\prod_{l \in C} V_{l}\right) =
\mbox{Re}\left(\prod_{l \in C} u_{l} \right)\quad,
  \label{wab}
\en

\noi where $u_{n,\mu} = \exp (i \theta_{n,\mu})$, $\theta_{n,\mu} =
\half \phi_{n,\mu}$. $C$ denotes a closed contour.

\noi In what follows, we will use $\theta_{n,\mu}$ to specify abelian
lattice fields for the sake of convenience.  The ``$\mbox{Re}$'' symbol
can be omitted from the definition of the abelian Wilson loop since
expectation values of this operator become automatically real due to
charge invariance ($\mbox{Im}\langle \prod_{l \in C} u_{l}\rangle=0$).

\subsection{Monopoles on the lattice}
One defines magnetic monopoles with respect to the residual $U(1)$
gauge group in the way proposed in Ref.~\cite{grantous} for $U(1)$
lattice gauge theory:  abelian plaquette variables,

\eq
\theta_{n,\mu\nu} =
\theta_{n,\mu} + \theta_{n+\hat{\mu},\nu} -
\theta_{n+\hat{\nu},\mu} - \theta_{n,\nu}\quad,\quad\theta_{n,\mu\nu} \in
[-4\pi,4\pi)\quad,    \label{206}
\en

\noi
can be decomposed into a periodic (regular)
part, $-\pi \leq \overline{\theta}_{n,\mu\nu} < \pi$, and a singular
part, $m_{n,\mu\nu} = 0, \pm 1, \pm 2$,

\eq
\theta_{n,\mu\nu} = \overline{\theta}_{n,\mu\nu} + 2\pi m_{n,\mu\nu}\quad.
\label{split}
\en

\noi $\overline{\theta}_{n,\mu\nu}$ describes the $U(1)$-invariant
``electromagnetic'' flux through the plaquette and $m_{n,\mu\nu}$ is the
number of Dirac strings passing through it.  Magnetic monopole
currents $k_{n,\mu}$, residing on the links of the dual lattice, are
defined as \eq k_{n,\mu} = \frac {1}{4\pi} \varepsilon_{\mu \nu \rho
\sigma} \partial_{\nu} \overline{\theta}_{n,\rho \sigma}\quad,
\label{208} \en \noi where a lattice forward derivative is used,
$\partial_{\nu} f_n = f_{n+ \hat{\nu}} - f_n $.  It is obvious, that
$k_{n,\mu}$ represents a conserved current:
\eq
\sum_{\mu} \partial_{\mu} k_{n,\mu} = 0\quad.
\en
Note, that the shifted
currents $j_{n,\mu}=k_{n+\hat{\mu},\mu}$ form closed loops on the dual
lattice.  As it has already been mentioned, first lattice results on
abelian dominance have been obtained in Ref.~\cite{suz1} where the
string tension computed from abelian Wilson loops
after MA projection,
Eq.~(\ref{wab}), has been numerically found to agree approximately
with the string tension as
extracted from the full, non-abelian theory.
Since then additional lattice results in support of
the 't Hooft-Mandelstam confinement scenario have been
found~\cite{suzrev,dgrev}. In particular the monopole
contribution to the abelian string tension has been investigated.

\subsection{Photon and monopole dynamics}
It is known that in the Villain formulation of compact $U(1)$ lattice
gauge theory one meets an exact factorization of expectation values of
Wilson loops into monopole and photon parts~\cite{banks} and one would
expect to encounter remnants of this  in other lattice formulations of
the theory, in the form
\eq \langle W^{U(1)}\rangle \approx \langle W^{mon}\rangle
\langle W^{ph}\rangle\quad.  \en
This factorization would induce a
decomposition of the potential between static charges at separation
${\mathbf R}$, \eq V^{U(1)}({\mathbf R}) \approx V^{mon}({\mathbf R}) +
V^{ph}({\mathbf R})\quad, \label{decomp1} \en where the photon
contribution is expected to be Coulomb-like.

 The decomposition rule, Eq.~(\ref{decomp1}), has indeed been verified
to hold approximately in simulations~\cite{stwen_u1} for the case of
compact $U(1)$ with Wilson action.  Similarly, in MA projected
$SU(2)$ gauge theory, abelian Wilson loop potentials were found to
decompose {\it qualitatively}~\cite{suz_monwl,stwen_su2}.

As we aim at more {\em quantitative} conclusions about the r\^ole of
monopoles and photons within gluodynamics we will proceed to
collect  all necessary formulae,  following essentially
Ref.~\cite{smitvds}.

The abelian Wilson loop operator Eq.~(\ref{wab}) can be easily
transformed to \eq W^{ab}(C) = \exp\left(- \frac{i}{2} \sum_{n, \mu
\nu} \overline{\theta}_{n,\mu\nu} M_{n,\mu\nu} \right)\quad,
\label{wab2} \en \noi where $M_{n,\mu\nu}$ is an integer valued
antisymmetric field, living on plaquettes, satisfying the condition, $
\partial^{-}_{\mu} M_{n,\mu\nu} = J_{n,\nu} $;  $J_{n,\nu}$ being
an external current associated to the (oriented) contour $C$ where it
takes the values $\pm 1$. $\partial^-_{\mu}$ denotes the lattice
backward derivative, $\partial^-_{\mu}f_n=f_n-f_{n-\hat{\mu}}$.
Let us now rewrite
$\overline{\theta}_{n,\mu\nu}$ in terms of dual potentials
$\rho_{n,\mu}$, and a photon field $\theta_{n,\mu}^{\prime}$,
\eq
\overline{\theta}_{n,\mu\nu} = \epsilon_{\mu\nu\alpha\beta}
\partial^{-}_{\alpha} \rho_{n,\beta} +
\partial_{\mu} \theta^{\prime}_{n,\nu} -
\partial_{\nu} \theta^{\prime}_{n,\mu}  + \overline{\theta}^0_{\mu\nu}\quad,
\label{dualp1}
\en
where
$\overline{\theta}^0_{\mu\nu}$ are zero modes defined by
\eq
\overline{\theta}^0_{\mu\nu} = \frac{1}{V} \sum_{n}
\overline{\theta}_{n,\mu\nu}
= - \frac{2 \pi}{V} \sum_{n} m_{n,\mu\nu}\quad.
\label{zeromo}
\en

\noi
The dual vector potential satisfies the equation,
\eq
\partial_{\nu} \partial_{\nu}^{-} \rho_{n,\mu} -
\partial_{\mu}^{-} \partial_{\nu} \rho_{n,\nu} = -2 \pi k_{n,\mu}.
\en
By imposing the Lorentz gauge condition ($\partial_{\mu} \rho_{n,\mu} = 0$)
one finds,

\eq
\rho_{n,\mu} = 2 \pi \sum_{m} D(n-m) k_{m,\mu} + \mbox{const.}\quad,
\label{dualp2}
\en
where $D(n-m)$ denotes the lattice Coulomb propagator.
Eqs.~(\ref{dualp1}), (\ref{zeromo}) and (\ref{dualp2}) define
$\theta'_{n,\mu}$ up to an irrelevant constant.
By inserting Eqs.~(\ref{dualp1}) and (\ref{dualp2}) into Eq.~(\ref{wab2}),
one  arrives at
\eq
 W^{ab} =   W^{mon}  W^{ph}
 W^{fv} \quad,
\label{factor}
\en
where
\eq
W^{mon} = \exp\left(-2\pi i\sum_{n,m} K_{n,\mu} D(n-m)
k_{m,\mu}\right)\quad,\quad
K_{n,\mu}= \half \epsilon_{\nu\mu\alpha\beta}
\partial_{\nu} M_{n,\alpha\beta}\quad,
\label{wlmon1}
\en
\eq
W^{ph} = \exp\left(-i \sum_{n} \partial_{\mu} \theta^{\prime}_{n,\nu}
M_{n,\mu\nu}\right)
= \exp\left(i \sum_{n} \theta^{\prime}_{n,\nu}  J_{n,\nu}\right)\quad,
\label{wlphot}
\en
\eq
W^{fv} =
\exp\left(-\frac{i}{2}
\overline{\theta}^0_{\mu\nu} \sum_{n} M_{n,\mu\nu}\right)\quad,
\label{wlfv}
\en
where the latter expression denotes a finite volume contribution.
After applying Eq.~(\ref{dualp1}), $W^{ph}$ can be written in a form,
that is more convenient for calculations,
\eq
W^{ph} =
\exp\left(-i \sum_{n,m} \partial^{-}_{\mu} \overline{\theta}_{n,\mu\nu}
D(n-m) J_{m,\nu}\right)\quad.
\en

It is worth mentioning that in Refs.~\cite{stwen_u1,suz_monwl,stwen_su2}
only monopole and photon contributions have been considered while the
finite volume contribution has been neglected.
 As pointed out above, for $U(1)$ gauge theory with Villain action,
it can be shown that Eq.~(\ref{factor}) holds for expectation
values of Wilson loops as well.

It is straightforward to generalize Eqs.~(\ref{wlmon1}) -- (\ref{wlfv})
to the case of ex\-tend\-ed mono\-poles~\cite{polik}
of size $l$. These are defined as

\eq
k^{(l)}_{n,\mu} =
\frac {1}{4\pi} \varepsilon_{\mu \nu \rho \sigma}
    \partial^{(l)}_{\nu} \overline{\theta}^{(l)}_{n,\rho \sigma}
    = \sum_{m \in c_l} k_{m,\mu}\quad,
\en
\noi where $\partial^{(l)}_{\nu} f_n = f_{n+ l \hat{\nu}} - f_n $,
$\overline{\theta}^{(l)}_{n,\mu \nu} = \sum_{m \in s_l}
\overline{\theta}_{n,\mu \nu}$, $c_l$ and $s_l$ are cubes and squares
of linear extent $l$. Subsequently, the operator
$W^{ab}(C)$ can be decomposed in the same way as in
Eq.~(\ref{factor}) with the monopole contribution defined by

\eq
W^{(l),mon} = \exp\left(-2 \pi i\sum_{n,m} K^{(l)}_{n,\mu} D^{(l)}(n-m)
k^{(l)}_{m,\mu}\right)\quad,\quad
K^{(l)}_{n,\mu}= \half \epsilon_{\nu\mu\alpha\beta} \partial^{(l)}_{\nu}
M_{n,\alpha\beta}\quad,
\label{wlmonl}
\en
\noindent $D^{(l)}(n-m)$ being the lattice Coulomb propagator on a lattice
of linear dimension $L/l$.

It is  useful to apply an alternative decomposition of the abelian
Wilson loop into monopole and photon parts, based on  the monopole
vector potential~\cite{smitvds}, $\theta^{mon}_{n,\mu} = \theta_{n,\mu} -
\theta^{\prime}_{n,\mu}$. From Eqs.~(\ref{split}) and (\ref{dualp1})
one  finds $\theta^{mon}_{n,\mu} $ to satisfy the condition
\eq
\partial_{\nu} \partial_{\nu}^{-} \theta^{mon}_{n,\mu} -
\partial_{\nu} \partial_{\mu}^{-} \theta^{mon}_{n,\nu} =
2 \pi \partial_{\rho}^{-} m_{n,\rho\mu}\quad,
\en
which has the Lorentz gauge solution,
\eq
\theta^{mon}_{n,\mu} = -2 \pi \sum_{m} D(n-m) \partial_{\nu}^{-}
m_{m,\nu\mu} + \mbox{const.}\quad.
\label{factor3}
\en

We can now write

\eq
W^{ab}(C) =  \exp\left(i \sum_{l \in C}\theta^{mon}_l  J_l\right)
\exp\left(i \sum_{l \in C} \theta^{\prime}_l  J_l\right)
\equiv W^{mon,fv}(C) W^{ph}(C)\quad.
\label{factor2}
\en
In Eq.~(\ref{factor2}), $W^{mon,fv}$ combines both, monopole and finite volume
contributions.
This representation has the advantage that --- provided the vector potentials
$\theta^{mon}_{n,\mu}$ and $\theta^{\prime}_{n,\mu}$  have been
computed --- smearing techniques~\cite{smearing} can be applied, which
have been proven
to be very useful in extracting potentials from Wilson loops.

\subsection{Charge two potential}
\label{charge2}
{}From Eq.~(\ref{coset}), the following decomposition of link variables
within the adjoint representation $U^A_{n,\mu}$ can be derived:
\eq
U^{A}_{n,\mu} = C^{A}_{n,\mu} V^{A}_{n,\mu}\quad ,
\en

\noi
where
\eq
(C^{A}_{n,\mu})_{\alpha \beta} = \half \tr\left[\sigma _{\alpha} C_{n,\mu}
\sigma _{\beta}  C^{\dagger}_{n,\mu}\right]
\en
and
\eq
(V^{A}_{n,\mu})_{\alpha \beta} = \half \tr\left[\sigma _{\alpha} V_{n,\mu}
\sigma _{\beta}  V^{\dagger}_{n,\mu}\right] =
\exp\left(i T_3 \phi_{n,\mu}\right)\quad.
\en
$T_3$ is the generator of the adjoint representation
and  the link angle $\phi_{n,\mu}$ has been defined above.

\noindent
The ``adjoint'' abelian Wilson loop,
\eq
W^{ab,adj}(C) = \frac{1}{3} \tr_{A}\left[\prod_{l \in C} V^{A}_{l}\right]\quad,
\en
can be expressed in terms of abelian link angles
\eq
W^{ab,adj}(C)  =  \frac{1}{3}(1+ 2 \cdot W^{ab,2}(C))\quad ,
\label{wladj}
\en
with
\eq
W^{ab,2}(C)  =  \cos(2 \theta_C)\quad,\quad
\theta_C = \sum_{l \in C} \theta_{l}\quad.
\en
The adjoint static ``quark'' field has one neutral and
two charged components with respect to residual abelian gauge
transformations.
The neutral component does not interact with
abelian gauge fields
and gives rise to the constant on the right hand side of
Eq.~(\ref{wladj}).
Thus, it is evident that, for a full description of the interaction
between adjoint sources, effects from the
off-diagonal gluon fields $C^{adj}_{n,\mu}$ have to be considered.
However, one still might expect abelian dominance to hold
for charged quarks.
We define the abelian projected analogue of the adjoint
potential, $V^{ab,adj}(R)$, from the asymptotic decay of large Wilson
loops $W^{ab,2}(R,T)$:
\eq
W^{ab,2}(R,T)
\propto \exp\left(-V^{ab,adj}(R)T\right)+\cdots\quad.
\en

We expect the string tension extracted from 
$V^{ab,adj}(R)$ to approximate the adjoint string tension
which can be defined at intermediate distances.
This expectation is supported
by an investigation of the abelian projection for the adjoint 
representation in the limit of large $N_c$~\cite{dfg1,dfg2}.

\section{Efficient gauge fixing}
\subsection{Overrelaxed simulated annealing algorithm}
As has been pointed out above the abelian projection procedure
requires (partial) gauge fixing.  The differential gauge condition
Eq.~(\ref{cgc}), supplemented by the constraint of Fadeev-Popov
operator positivity, is equivalent to finding a maximum of the lattice
functional, Eq.~(\ref{maxfunc}).  Besides absolute maxima --- which
can in principle be degenerate even beyond trivial gauge
transformations, such as constant ones and transformations within the
maximally abelian subgroup --- the functional $F(U)$ can have any number
of local maxima.  This feature resembles the Gribov problem of
continuum gauge theories~\cite{grib}.  While degenerate absolute
maxima, at least for Landau gauge, may be safely ignored~\cite{zwanziger},
imperfect gauging (where the system is stuck
on some local maximum) may lead to fake physics results~\cite{bmmp}.
The aim of any reliable analysis must therefore be to drive the system
by appropriate gauge procedures into local maxima as close as possible
to the absolute ones.  This would help to  reduce systematic
uncertainties due to gauge fixing ambiguities.
A similar approach of gauge fixing in numerical simulations,
motivated by global gauge fixing conditions~\cite{parjl,zwanz2},
has been advocated for Landau and Coulomb gauges in Ref.~\cite{parfac}.

Traditionally, the relaxation plus overrelaxation (RO) algorithm has
been employed for MA gauge fixing. In Ref.~\cite{bbmp}, we reported on
the implementation of simulated annealing (SA), a
technique that has been
proven to be very useful in handling various optimization
problems. In this SA algorithm~\cite{kgv,cerny}
the functional $F(U)$ is regarded as a ``spin action'',

 \eq S(s)=F(U^g) =
\frac{1}{8 V} \sum_{n,\mu} \tr\left( s_n U_{n,\mu} s_{n+ \mu}
U_{n,\mu}^{\dagger} \right) \en
\noindent
where $s_n=g_{n}^{\dagger} \sigma_3 g_{n}$ resemble spin variables.
The lattice fields $U_{n,\mu}$ play the r\^ole of (almost) random
local couplings.
Maximizing the functional $F(U^g)$ is equivalent to decreasing the
auxiliary temperature $T$ of the statistical system with
partition function
\eq
  Z = \sum_{\{s_n\}} \exp\left(\frac{1}{T}S(s)\right)\quad.
\en

One starts with equilibrating this spin glass at high temperature.
Subsequently, $T$ is decreased adiabatically. It is evident that in the limit
$T \ra 0$ the system approaches its ground state, i.e.\ the maximal
value of $S$.  The merits of SA can be phrased in the language of
solid state physics: standard relaxation that corresponds to fast
cooling
might cause defects which are avoided in the adiabatic
cooling procedure of SA.  In order to enhance  the adiabatic
movement of the spin variables through phase space we complement
simulated annealing with overrelaxation (OSA).

Our procedure consists of three steps:
\begin{enumerate}
\item Thermalization at $T=2.5$.
\item Gradual decreasing of $T$ down to $T=0.01$.
\item Final maximization by means of the RO algorithm.
\end{enumerate}

In steps 1 and 2 an overrelaxation transformation is performed at six
consecutive lattice sites and heatbath is applied to the seventh.
Within step 2, every time when the heatbath update is applied to a
site, the temperature is lowered by a quantum $\delta T$.  For our
large volume studies ($V=32^4$), a variant of this algorithm (suitable
for parallel computers like the Connection Machine CM-5 where sites
are visited in lexicographical ordering within subcubes of $2^4$ sites
each) has been employed.  The combined effect of local overrelaxation
plus local temperature reduction is to cut  the number of cooling
sweeps while remaining close to equilibrium.

For the initial thermalization at $T=2.5$, 20 sweeps have been performed.
Within the temperature range $2.5\geq T\geq 0.1$,
$\delta T (T)$ has been tuned such that the spin action
increased about linearly with the number of iteration sweeps.
This has been realized by subdividing this range into 24 intervals of
width $\Delta T=0.1$.

The corresponding differences of the action $\Delta
S(T)=S(T)-S(T-\Delta T)$ have been computed on equilibrated
configurations and were
 found to be very stable against statistical
fluctuations among different Monte Carlo (MC) configurations. We found
practically no
volume dependence and only a moderate impact from variations
of the gauge coupling, $\beta$.
The number of sweeps (out of a fixed total number)
to be performed within each interval $(T-\Delta T,T]$ was chosen to
be proportional to $\Delta S(T)$ and, subsequently, the corresponding
value of $\delta T(T)$ has been determined.
Within the region $0.1>
T\geq 0.01$, 50 additional sweeps have been performed. Finally, the RO
algorithm has been applied till a convergence criterion was
satisfied\footnote{The iterations have been stopped as soon as all
rotations of $s_n$ among the lattice sites within a maximization sweep
were equal to identity within single precision numerical accuracy.}.

\subsection{Tuning the gauge fixing algorithm}
Prior to the large volume production runs, we compared the RO gauge
fixing algorithm with three different OSA variants, using a $12^4$
lattice at $\beta = 2.43$.  The OSA variants employ three different
cooling schedules at step 2 of our procedure: 250 (OSA1), 500 (OSA2)
and 1000 (OSA3) total sweeps.  The standard procedure (RO) has been
applied with identical convergence criterion. We collected 30
statistically independent equilibrated MC configurations and produced
10 random gauge copies from each of them as inputs for the four
algorithms.

Consider a gauge dependent abelian quantity $A$.  In the following, we
denote $\overline{A}$ to be the average over gauge copies and $\langle
A \rangle$ to be the statistical average.  In Table~\ref{t1}, we compare the
output of the four algorithms for various quantities.
$\delta_{sd}^2=\langle \overline{F^2}-\overline{F}^2\rangle$ denotes
the scatter of the maximized value of the functional among gauge
copies. An ideal algorithm would always yield $F=F_{\max}$, i.e.\
$\delta_{sd} =0$.  $\rho_{\mbox{\scriptsize
mon}}=1/(4V)\langle\sum_{n,\mu}k_{n,\mu}\rangle$ denotes the
so-called monopole
density.  $W_{ij}$ are abelian $i\times j$ Wilson loops and $K^{ab}$
is the abelian string tension in lattice units.

 The comparison of $\langle \overline{F}\rangle$ and $\delta_{sd}$
reveals that anyone of our OSA schedules is superior to the standard
RO algorithm, with the longest schedule yielding the best
results. The  OSA2 and OSA3 algorithms perform equally  on the
level of statistical errors.  In terms of total computer time spent
with a scalar code, OSA2 is about a factor two slower than RO.  OSA1,
still being an improvement, is only 30 \% slower than RO.  In our
parallel implementation on a Connection Machine CM-5, OSA1 was found
to run  even slightly faster than RO.

The lesson from Table~\ref{t1} is that physical results can be substantially
corrupted by inadequate gauge fixing. The functional $F$ turns out to
be correlated with Wilson loops and anti-correlated with the monopole
number and abelian string tension.

\begin{figure}[p]
\begin{center}
\leavevmode
\hbox{
\epsfxsize=14truecm\epsfbox{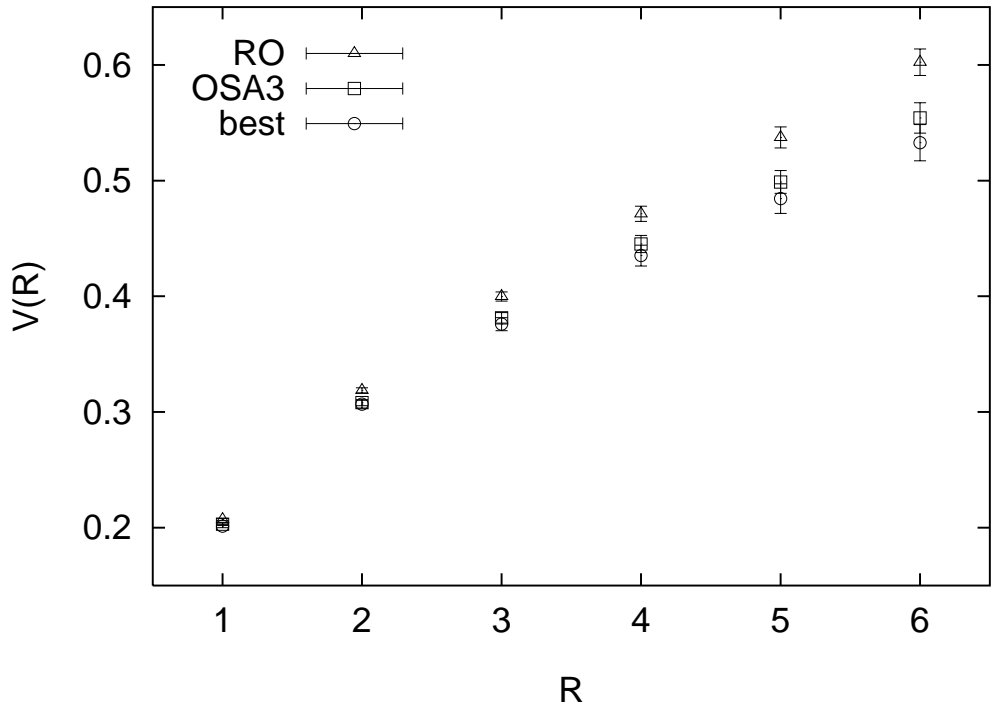}
     }
\end{center}
 \caption{Abelian potentials from RO (triangles), OSA3 (squares)
 and ``best'' copy (circles). }
\label{fig:sa_poten}
\end{figure}

By applying state-of-the-art smearing techniques on the spatial
transporters of abelian Wilson loops, we have been able to
compute the abelian ground state potential.
In Fig.~\ref{fig:sa_poten} these potentials are displayed
for the RO (used by all previous authors for MA gauge fixing)
and the OSA3 algorithms
with only one gauge copy used in both cases.
In addition, the copy with largest functional among all
OSA3 copies has been chosen on each configuration as
our best estimate of the ``true''
maximum (circles).
{}From this potential we obtain $K^{ab}_{\mbox{\scriptsize
best}} = .0478(38)$ as an
estimate of the abelian string tension\footnote{Within this section,
all errors have been
obtained by the jackknife procedure.}.
{}From Fig.~\ref{fig:sa_poten}
it is evident that this value will be overestimated by
about $30\%$ by use of the standard procedure (see also Table~\ref{t1}).
Even our
most expensive
algorithm OSA3 yields a value that is off by about one
statistical standard deviation
{}from our best estimate
($K^{ab}_{\mbox{\scriptsize OSA3}} = .0536(30)$).

In order to study the scaling properties, the investigations have been
partly repeated on a lattice with nearly equal physical size but
smaller lattice resolution ($16^4$ at $\beta=2.5115$).  The OSA2 and
RO algorithms have been applied to 10 copies on 20 configurations,
each.  The qualitative properties are seen to be the same.  Again, the
string tension obtained after application of the RO algorithm is
drastically  overestimated: $K^{ab}_{\mbox{\scriptsize RO}} =
.0362(15)$ versus $K^{ab}_{\mbox{\scriptsize best}} = .0305(19)$.
Since the non-abelian string tension at this $\beta$ value turned out
to be $K=.0325(12)$, this difference is relevant to the physical
conclusions drawn.

We conclude that the quality of MA gauge fixing (in terms of the value
of $F(U^g)$ and the scatter of results among different gauge copies)
can be significantly improved by applying the OSA algorithm, without
any cost in computer time.  Our systematic
study of non gauge invariant quantities revealed that such improvement is
in fact mandatory: for a reliable extraction of the abelian potential
and other observables one must carefully eliminate biases from
incomplete gauge fixing.
An extension of the method to other gauges (e.g.\
Landau gauge) is straight forward.

\section{Biases from Gribov copies}
\subsection{Simulation technicalities}
\noindent
Our main simulations have been performed on $32^4$ lattices at
$\beta=2.5115$.  As a first step, test runs with different OSA cooling
schedules (as explained in the previous section) have been performed
on four gauge copies, generated from two thermalized configurations
with the above simulation parameters.  We have chosen the following
schedules for the temperature degrading sweeps: $N_s=250+50$,
$N_s=500+50$, $N_s=1000+100$, $N_s=2000+100$, and $N_s=5000+200$. The
first numbers denote the sweeps spent within the linear region
($0.1<T<2.5$). The latter numbers correspond to the sweeps applied
within the interval $0.01<T<0.1$.  Finally, direct maximization sweeps
($T=0$) have been applied.  Depending on the configuration and cooling
schedule typically 10--100 such steps had to be performed until the
rest vector criterion was satisfied.

\begin{figure}[p]
\begin{center}
\leavevmode
\epsfxsize=14truecm\epsfbox{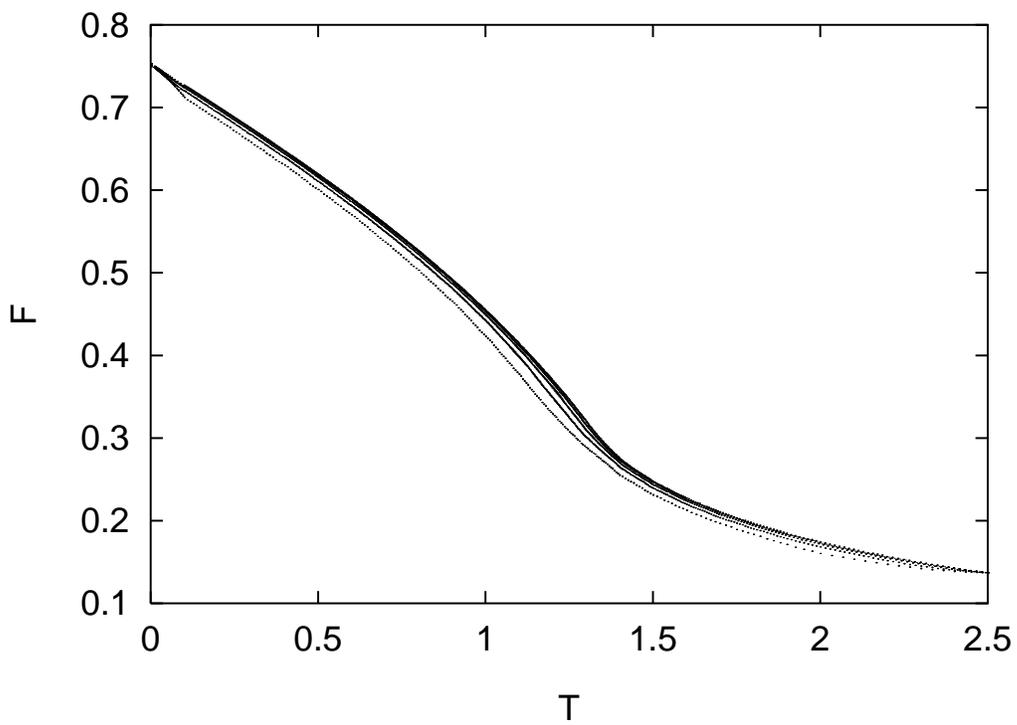}
\end{center}
\caption
{
The spin functional $F$ as a function of the temperature $T$
for various cooling schedules. The curves
correspond to $N_s=5200$, $N_s=2100$, $N_s=1100$,
$N_s=550$, and $N_s=300$, respectively.}
\label{fig:hyst}
\end{figure}

The hysteresis curves of the spin action, $F$, as a function of the
temperature, $T$, are displayed in Fig.~\ref{fig:hyst}.  The upmost
curve corresponds to the longest schedule. The scatter
within each curve (which is not visible on the scale of the figure)
is due to the four gauge copies generated
and indicative for the (small) statistical uncertainty.
The differences between
the curves indicate that even for $N_s>2000$ thermal equilibrium is
not yet reached.

For our final run, we decided to apply 1100 gauge fixing sweeps as a
compromise between effort and outcome.  This choice allows to
generate a few local maxima of $F(U)$ on each configuration and
subsequently to select the best out of them. As will be described
below, the availability of several  local maxima  enables us to
estimate systematic biases due to incomplete gauge fixing.

\subsection{Error estimates}

\noindent
In order to estimate systematic errors induced by an incomplete gauge
fixing procedure, we have generated $N=20$ random gauge copies on each
of $N_c=30$ gauge configurations. Subsequently, these copies have been
fixed to the MA gauge.
Each abelian configuration (i.e. each gauge
copy) $C_j(i)=\{\theta_{\mu}^{(j,i)}(n):\mu=1,\ldots,4, n\in V\}$
is labeled in the following way: $j$ runs from one to the number of
gauge copies, $N$, while $i$ runs from 1 to the number of
Monte Carlo generated $SU(2)$ gauge
configurations, $N_c$. On each of these copies, abelian
quantities $A(C_j(i))$ are measured, where $A$ denotes either
a (smeared) Wilson loop, the plaquette or the monopole number. In addition,
the values of the gauge fixing functional $F(C_j(i))$ are stored.
All gauge copies (on a given configuration) are sorted by the value of
this functional:
\begin{equation}
F(C_1)\leq F(C_2)\leq\cdots\leq F(C_N) \quad.
\end{equation}

\begin{figure}[p]
\begin{center}
\leavevmode
\epsfxsize=14truecm\epsfbox{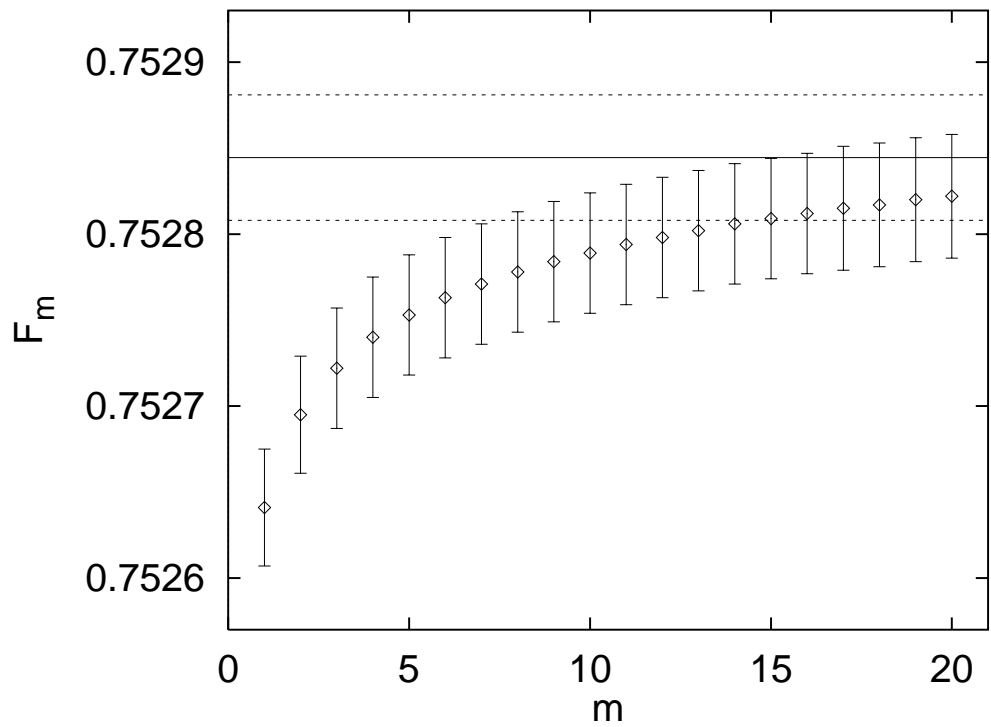}
\end{center}
\caption
{The maximized value of the spin functional $F$ as a function
of the number of local maxima generated on each configuration.
The solid line (with error band) represents our extrapolated value
($m\rightarrow\infty$).}
\label{fig:func}
\end{figure}

We are now prepared to investigate, which value of $A$ we would have
obtained on the ``best'' out of $m\leq N$ copies. To answer this
question one has to select $m$ random copies out of the $N$ copies
that have been generated in total and subsequently  extract $A$ from
the copy with largest functional $F$, i.e.\ largest index $i$.
Averaging over all possible choices yields
\begin{equation}
\label{eqcombo}
A_m=\left(\begin{array}{c}N\\m\end{array}\right)^{-1}\times\sum_{j=m}^N
\left(\begin{array}{c}j-1\\j-m\end{array}\right)A(C_j)
\end{equation}
as the ``average best copy'' expectation of $A$
on subsets of size $m$ where the gauge configuration index has been
omitted. As expected, the above formula corresponds to the average
over all copies for the special case $m=1$ while for $m=N$ one obtains
the value of $A$ on the ``best'' copy.
Quantities like the abelian potential can be
computed from averages of these $A_m$'s over the $N_c$
configurations.
The situation is visualized in Fig.~\ref{fig:func} for the gauge
functional $F$ itself.
Throughout the rest of this paper, all statistical errors have been
computed by the bootstrap procedure.

We wish to determine the expected deviation of the estimate $\langle
A_m\rangle= \frac{1}{N_c}\sum_{i=1}^{N_c}A_m(i)$ from the true value
$\langle A_{\infty}\rangle$, related to an absolute maximum.  To
estimate this bias  we make the following assumptions:
\begin{itemize}
\item The absolute maximum\footnote{The term ``absolute maximum''
is to be understood
modulo trivial degeneracies, due to constant gauge
transformations, gauge transformations within the unfixed diagonal
$U(1)$ subgroup and the $2^4$ degenerate maxima from $Z_2$ center group
transformations within hyperplanes perpendicular
to the four possible lattice orientations, which only affect
expectation values of Polyakov line-like operators but leave the
spectrum invariant.}
is unique.
\item The algorithm is in principle able to reach the absolute
maximum $F_{\mbox{\scriptsize global}}$, i.e.\ $F_{\infty}=
F_{\mbox{\scriptsize global}}$.
\item In the neighborhood of the absolute maximum, $\langle A_m\rangle$
approaches $\langle A_{\infty}\rangle$ as a mono\-ton\-ous function of
$\langle F_{\infty}-F_m\rangle$.
\end{itemize}
The first assumption is supported by Zwanziger's proof of the
non-degeneracy of absolute maxima within the interior of the
fundamental modular
region in Landau gauge
on the lattice~\cite{zwanziger} while the third assumption is
supported by numerical evidence as all our observables exhibit strong
correlations with the average value of the functional.

\begin{figure}[p]
\begin{center}
\leavevmode
\epsfxsize=14truecm\epsfbox{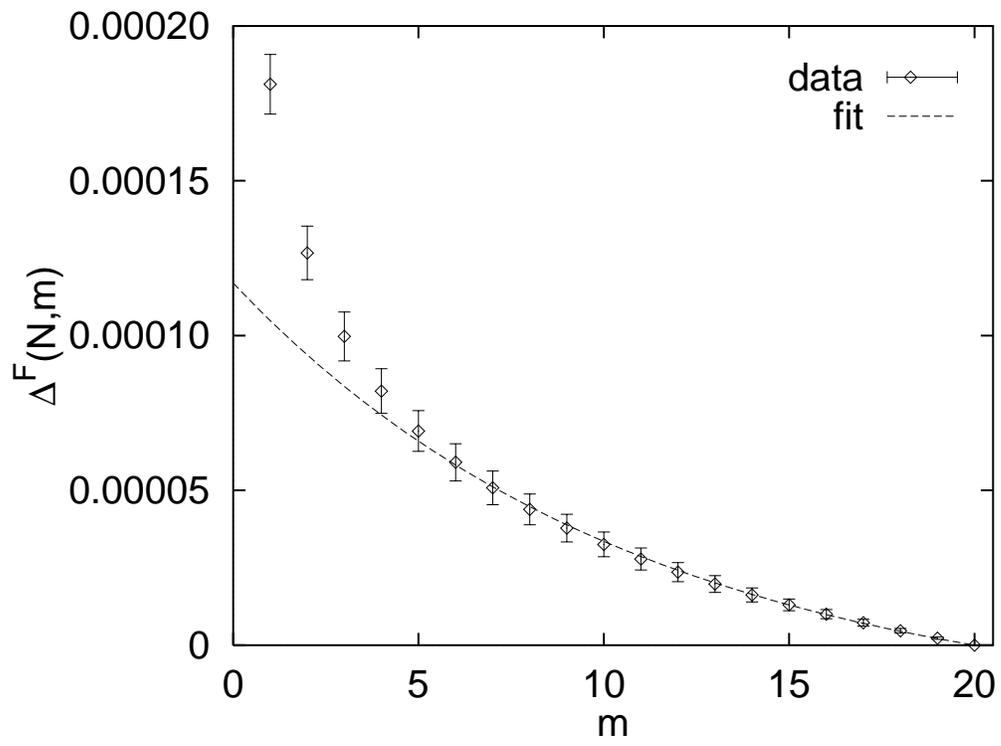}
\end{center}
\caption
{The differences, $\Delta^F(N,m)=F_N-F_m$ and an exponential fit.}
\label{fig:func1}
\end{figure}

The difference, $\Delta_m^A=\langle A_{\infty}-A_m\rangle$, is to be
seen as the very bias on $A_m$ from incomplete gauge fixing.  The
statistical uncertainty on this bias, on the other hand, is nothing
else but the systematic error on our final result. As to its
$m$-dependence, we start from the
ansatz
\begin{equation}
\Delta_m^A=c_1\exp(-d_1m)+\cdots
\end{equation}
for the (large $m$) asymptotic behavior, to be tested against the
data. Accordingly, we can fit our data to the form
\begin{equation}
\langle A_m\rangle =\langle A_\infty\rangle-c_1\exp(-d_1m)+\cdots
\label{fitti}
\end{equation}
where $\langle A_\infty\rangle$, $c_1$ and $d_1$ are free parameters.
The statistical errors on
$\langle A_m\rangle$ imply a large uncertainty on $\langle A_\infty\rangle$.
However, due to strong
correlations among the data, the differences
$\Delta_m^A$
can be obtained quite accurately. In view  of our limited statistics
for the study of gauge fixing ambiguities
($N_c=30$) we have not applied full correlated fits.
Nonetheless, some of the correlations have been taken into account by
fitting the differences
\begin{equation}
\Delta^A(N,m)=
\langle A_N-A_m\rangle=\Delta^A_N\left(e^{(N-m)d_1}-1\right)\quad,\quad
\Delta^A_N=c_1e^{-Nd_1}
\end{equation}
rather than following Eq.~\ref{fitti}.  The result of such a fit
(from $m=5$ onwards) is visualized in Fig.~\ref{fig:func1}. On the
basis of this fit the bias on $A_m$ can be traced back into the region
of small $m$, in the form $\Delta^A_m=\Delta^A_N+\Delta^A(N,m).$

Results from the above procedure for the functional, $F$, the abelian
plaquette action, $S=1-\langle W^{ab}(1,1)\rangle$, and the monopole
density $\rho_{\mbox{\scriptsize mon}}$ for $m=1, 10, 20$ as well as
for the extrapolated value ($m= \infty$) are compiled in
Table~\ref{t2}.
We find that the proposed error analysis is a powerful tool to obtain
reliable estimates for biases. For instance, when selecting the best
out of 10 copies generated by our OSA algorithm, we find $\Delta
F_{10}=0.000055(10)$, $\Delta S_{10}=-0.000037(30)$, and
$\Delta\rho_{\mbox{\scriptsize mon},10} = -0.000015(15)$.

We would like to emphasize that, by computing the biases, we have
found a way to extrapolate values, obtained on local maxima, to an
absolute maximum. The accuracy of all computations is limited by the
statistical error on the biases.

\subsection{Application to the abelian potential}
\noindent
The abelian potential, $V^{ab}_m({\mathbf R})$, as well as the abelian string
tension, $K^{ab}_m$, have been computed for various $m\leq N$,
by use of the method described in Section~5.1. We find the abelian string
tension, $K^{ab}_m$ to be anticorrelated with the gauge fixing functional
($K^{ab}_m<K^{ab}_{m-1}$). The potential values themselves
exhibit a systematic, but statistically insignificant, drift.

\begin{figure}[p]
\begin{center}
\leavevmode
\epsfxsize=14truecm\epsfbox{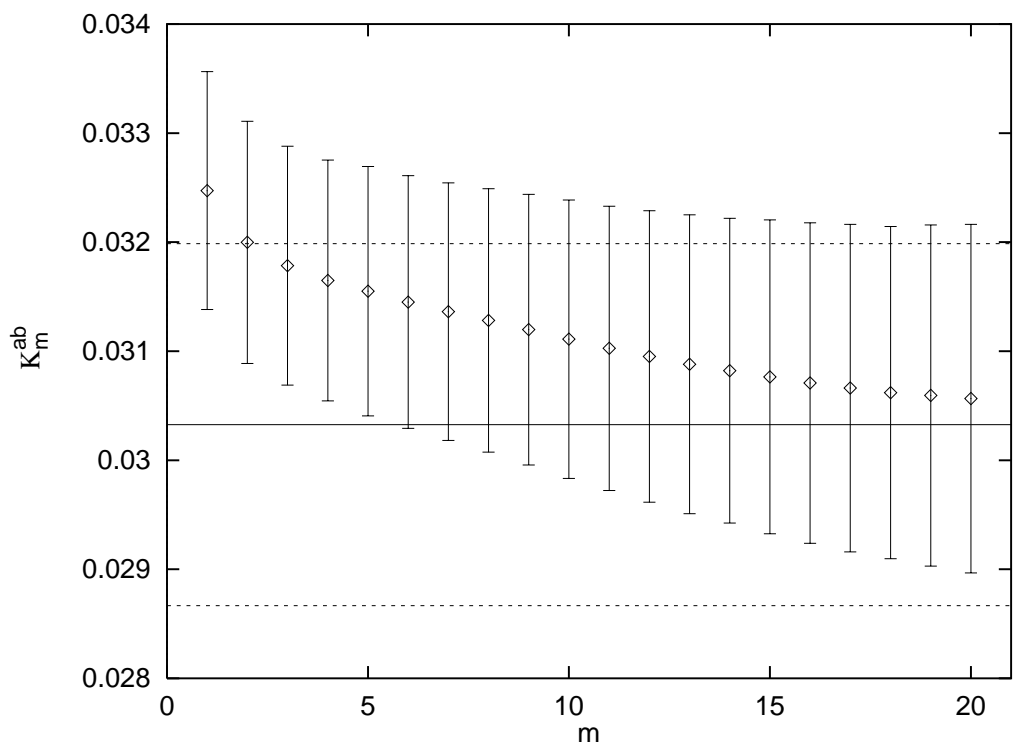}
\end{center}
\caption
{The abelian string tension versus the number of MA gauge copies
$m$.
The solid line (with error band) represents our extrapolated value
($m\rightarrow\infty$).}
\label{fig:kappa}
\end{figure}

\begin{figure}[p]
\begin{center}
\leavevmode
\epsfxsize=14truecm\epsfbox{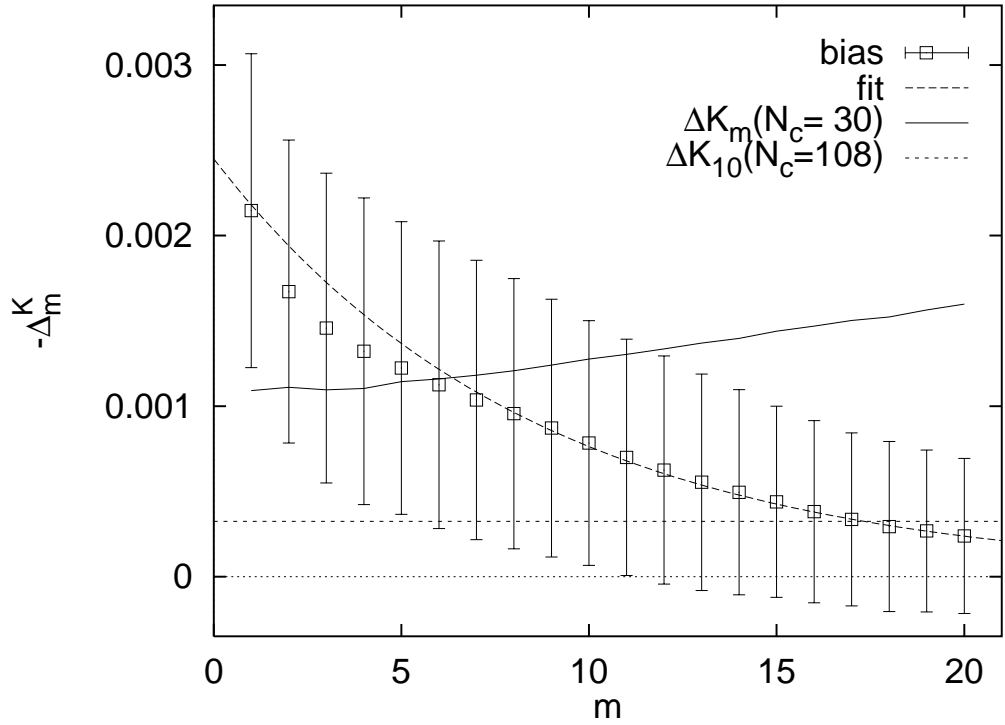}
\end{center}
\caption
{Differences between the extrapolated value $K^{ab}_{\infty}$ (from the
exponential fit) and $K^{ab}$, obtained on a finite number of gauge
copies, $m$. The solid line denotes the
statistical uncertainty on $K^{ab}$ from
30 gauge configurations. The horizontal
dashed line is the statistical
uncertainty on 108 configurations with $m=10$.}
\label{fig:kappa2}
\end{figure}

The correlation between the fitted abelian string tension and $m$ can
be read off from Fig.~\ref{fig:kappa}. In analogy to the extrapolation
method discussed in the previous section, $\Delta^K(N,m)$ is fitted to
an exponential ansatz.  As a result, the systematic bias,
$\Delta^K_m=\Delta^K_N-\Delta^K(N,m)$ is obtained as a function of
$m$ (Fig.~\ref{fig:kappa2}).  The statistical errors on $N_c=30$
configurations\footnote{The statistical errors come out to decrease
with decreasing $m$ due to the procedure of combinatoric averaging,
Eq.~\ref{eqcombo}.} (solid line)
are indicated in the figure, as well as the error on the final
statistical ensemble of $108$ configurations, obtained on the best out
of 10 gauge copies (horizontal dashed line).

For the final production runs  we decided to choose $m=10$ as a reasonable
compromise. For this $m$-value the expected statistical error on the
string tension matches both, the size and the uncertainty of its bias.

Contrary to the case of the string tension (Fig.~\ref{fig:kappa2}), we
find no statistically significant bias on the values of the
Coulomb coefficient $e$
and the self energy $V_0$ which are dominated by the short range part
of the potential.

Results on the abelian string tension from different numbers of gauge
copies are collected in Table~\ref{t2} (last row). The bias on 10 gauge
copies is $\Delta^K_{10}=-0.00078(72)$ while the expected bias on one
gauge copy would have been $\Delta^K_1=-0.0021(9)$.

\section{Physics results}

Our main measurements have been performed on the same lattice volume
and $\beta$ value as the investigation of systematic gauge fixing
errors, presented in the previous section ($V=32^4$, $\beta=2.5115$).
This enables us (a) to correct the results for the estimated biases
and (b) to include systematic uncertainties into the final errors.  The
spatial lattice extent corresponds to 2.7~fm in physical units where
the scale has been obtained from the value
$\kappa=Ka^{-2}=(440\,\,\mbox{MeV})^2$ for the string tension.  This
ensures finite size effects on the potentials to be negligible. The
$\beta$ value was chosen sufficiently large to be within the scaling
region, and in respect to future finite temperature studies (The
critical temperature comes out to be $T_c=(8a)^{-1}$~\cite{fingb}).

The hybrid-overrelaxed algorithm~\cite{hor}
with Fabricius-Haan heatbath sweeps~\cite{fab} has been applied to update
the gauge fields. Subsequent configurations are separated by 200 such
sweeps and have been found to be statistically independent.
Abelian projection and measurements have been performed on
108 such gauge configurations. The following abelian observables
(i.e. quantities expressed in terms of abelian gauge fields,
$\theta_{n,\mu}$) have been investigated:
charge one and two potentials and string tensions, photon
and monopole contributions to the abelian string tension and the abelian
monopole density.

\subsection{Abelian and non-abelian static potentials}
We have computed the abelian potential on 108 configurations as well
as the non-abelian potential on 644 configurations by use of one and
the same analysis method to allow for direct comparison of
results.

Our final results on the abelian potential have been obtained
on the best out of $10$ gauge copies. The systematic biases are
estimated from 30 configurations with 20 gauge copies on each.  Both,
abelian and non-abelian potentials have been obtained from Wilson
loops with smeared spatial paths in order to enhance the overlap of
the $Q\overline{Q}$ creation operator with the $Q\overline{Q}$ ground
state~\cite{smearing}. To reduce statistical fluctuations, we have
analytically integrated out temporal links of the non-abelian Wilson
loops~\cite{linkint}.

We found good results by iteratively applying the
smearing procedure,
\eq
\theta_{n,j}\ra\arg\left[\alpha \exp\left(i\theta_{n,j}\right)+\sum_{k\neq j}
\exp\left(i\left(\theta_{n,k}+\theta_{n+\hat{k},j}-\theta_{n+\hat{j},k}\right)
\right)\right]\quad,
\en
to spatial link angles, i.e.\ by substituting the corresponding $U(1)$
element by a linear combination of the previous one and the sum of
the four spatial staples, enclosing it.
We have chosen the parameter value
$\alpha=1$ and 150 iterations. Rectangular
Wilson loops, constructed from such smeared
spatial links, can be decomposed into a linear combination of various
loops with fixed corners but different spatial connections. For
extraction of the non-abelian potential, we applied a similar procedure
on the $SU(2)$ link variables
with the parameter value $\alpha=2$.

For large temporal extent, $T$,
the potential can be extracted from the asymptotic expectation,
\begin{equation}
\langle W({\mathbf R},T)\rangle
=C({\mathbf R})\exp\left(-V({\mathbf R})T\right)+\cdots\quad,
\end{equation}
where $C({\mathbf R})$ denote the ground state overlaps.
For finite values of $T$, we define the following approximants to
overlaps and potentials, which will monotonously decrease towards their
asymptotic values,
\begin{equation}
V_T({\mathbf R})=\log\left(\frac{\langle W({\mathbf R},T)\rangle}
{\langle W({\mathbf R},T+1)\rangle}\right)
\quad,\quad
C_T({\mathbf R})=\langle W({\mathbf R},T)\rangle
\exp\left(V_T({\mathbf R})T\right)\quad.
\end{equation}

By comparing results on the potential approximants (abelian and
non-abelian) between the on-axis direction with those, obtained on
five different off-axis directions, we find $SO(3)$ rotational
invariance to be restored (within statistical accuracy) for $R\geq
3$. In both cases, the data for $R\geq 2\sqrt{3}$ are well described
(for $T\geq 3$) by the parametrization,
\begin{equation}
\label{potffit}
V(R)=V_0+KR-\frac{e}{R}\quad.
\end{equation}
By fitting different $T$-approximants to the potential with this
parametrization, we obtain approximants to the string tension, $K_T$.
By demanding plateaus of $C_T(R)$, $V_T(R)$ and $K_T$ for $T\geq
T_{\min}$, we find $T_{\min}=4$ and $T_{\min}=6$ for the abelian and
non-abelian potentials, respectively. The different onset of
asymptotics is due to superior overlaps with the ground state in case
of the smeared abelian Wilson loops.  The self energy
term, $V^{ab}_0$ is much smaller than its abelian counterpart $V_0$,
resulting in larger numerical values of the corresponding abelian
Wilson loops. These two effects are among the reasons for
reduced statistical errors on the
abelian potential and fit parameters.  We take the $T_{\min}$
approximants to potential values and fit parameters as our asymptotic
results. To avoid systematic effects from the fit range creeping into
the comparative interpretation of results, we select a universal
$R$-range for all fits,  $2\sqrt{3}\leq R \leq 16$. This provides us
with 45 on- and off-axis data points.

\begin{figure}[p]
\begin{center}
\leavevmode
\epsfxsize=14truecm\epsfbox{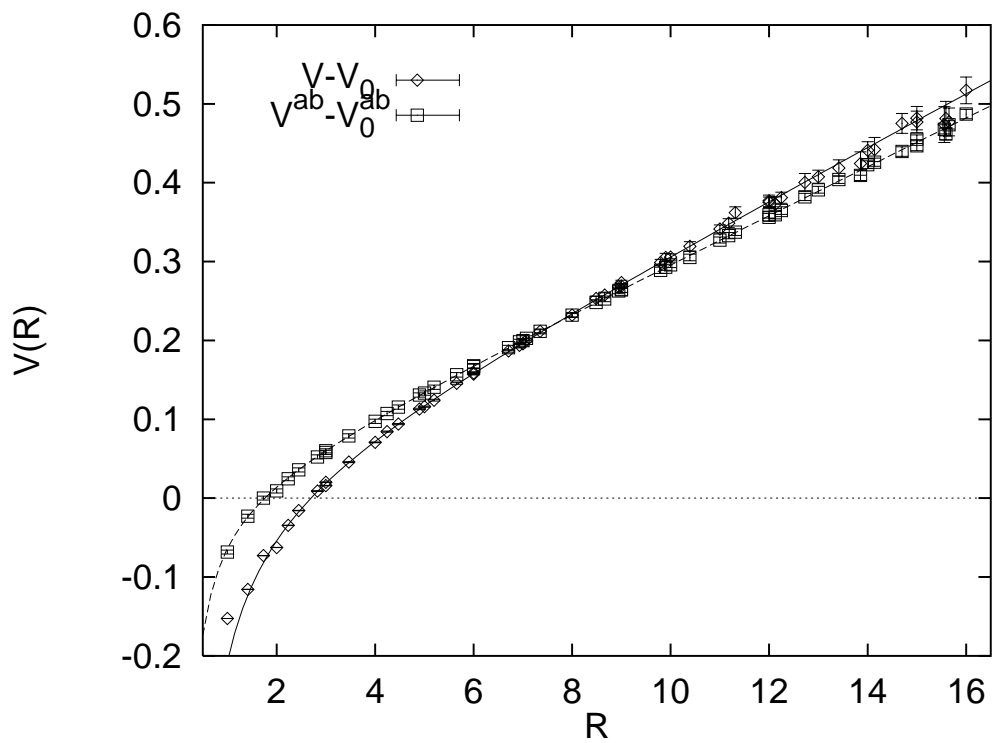}
\end{center}
 \caption{The abelian and non-abelian potentials, $V^{ab}$ and $V$. }
\label{fig:ab_nab}
\end{figure}

Our results on the fit parameters are in {\em qualitative} agreement with
previous publications \cite{suzrev}.  In particular, the self energy
$V^{ab}_0$ and Coulomb coefficient $e^{ab}$ come out to be by more
than a factor two smaller than their non-abelian counterparts while
the abelian string tension is found to be close to the non-abelian
one (see Table~\ref{t3} and Fig.~\ref{fig:ab_nab}).  As pointed out above,
all systematic uncertainties are understood and under control in the
present investigation, in particular the approach to the
$T\rightarrow\infty$ limit. The biases due to gauge fixing ambiguities
have been neglected in previous studies but turn out to be
important as demonstrated in Sections 3 and 4.

The value of the abelian string tension comes out to be $K^{ab}=
0.0305(3)$ on the best out of 10 OSA gauge copies, which nicely agrees
with the value $K^{ab}=0.0305(19)$, as obtained on a $16^4$
lattice (Section 3) at the same $\beta$ value.  By including bias and
uncertainty of the bias, we end up with $K^{ab}=0.0297(8)$, where the
error includes the systematic uncertainty.  This amounts to the ratio
$K^{ab}/K=0.92(4)$ (see Table~\ref{t3}).

Given the precision of our data and
analysis tools, we find a significant deviation between the string
tension of $SU(2)$ gauge theory
and its MA content at $\beta = 2.5115$.  Further
measurements at different lattice spacings are required to decide
whether this ratio approaches unity in the continuum limit, as
expected from abelian dominance. However, for the time being,
at finite lattice
spacing, the string tension from the abelian projected theory appears
to be definitely smaller than its non-abelian counterpart.

\subsection{Decomposition of the abelian potential}

We have tried two different approaches to disentangle monopole and
photon contributions to the potential.

The first method rests on the determination of the monopole part from
elementary monopole currents, $k_{n,\mu}$. The potential estimators
$V^{mon}_T(R)$, extracted from $\langle W_{mon}(R,T)\rangle$ (see
Eq.~(\ref{factor})) at fixed $R$ have been observed to increase with
$T$, which means that the coefficients $C'(R)$ of the decomposition,
\eq \langle W^{mon}(R,T)\rangle = C(R) \exp\left(-V^{mon}(R) T\right)
+ C'(R) \exp\left(-V'{}^{mon}(R) T\right) + \cdots\quad, \en are not
necessarily positive. This unpleasant feature, in conjunction with the
requirement of $T\gg R$ makes this procedure of extracting
$V^{mon}(R)$ unreliable.

As a way out, we started from an alternative representation of the
factorization property to $W^{ab}$ (Eqs.~(\ref{factor3}) and
(\ref{factor2})), which allows decoupling of excited states through
smearing of spatial links $\theta^{mon}_{n,j}$.
In this second approach , the abelian configuration has been fixed for
technical reasons to Landau gauge,
\eq
\partial_{\mu} \sin(\theta_{n,\mu}) =0\quad,
\en
prior to evaluation of $\theta^{mon}_{n,\mu}$ (Eq.~(\ref{factor3})).
This reduces the number of plaquettes with nonzero $m_{n,\mu\nu}$ and,
thus, computer time by more than one order of magnitude.  It should be
noted that --- though $\theta^{mon}_{n,\mu}$ transforms under $U(1)$ gauge
transformations Eq.~(\ref{abgtr}) --- $W^{mon,fv}$ as well as $W^{ph}$
are gauge invariant quantities.

After evaluation of $\theta^{mon}_{n,\mu}$ and $\theta'_{n,\mu} =
\theta_{n,\mu} - \theta^{mon}_{n,\mu}$, smearing has been applied and
the corresponding potentials have been extracted. For small $R$, the
results as obtained from the two methods are found to agree, while at
large $R$, plateaus in $T$ could only be established for potential
estimates extracted from smeared monopole and photon Wilson loops.

As pointed out in Section~2, the full factorization ansatz may include
finite volume contributions.  However, no statistically significant
such effects have been found by comparison of unsmeared $W^{mon,fv}$
with $W^{mon}$ (calculated by use of elementary monopole currents
$k_{n, \mu}$ via Eq.~(\ref{factor}) for all realized $R$ and $T$
values, on a configuration by configuration basis).  Thus, we conclude
that finite size effects can indeed be neglected on our lattice
volume.

\begin{figure}[p]
\begin{center}
\leavevmode
\epsfxsize=14truecm\epsfbox{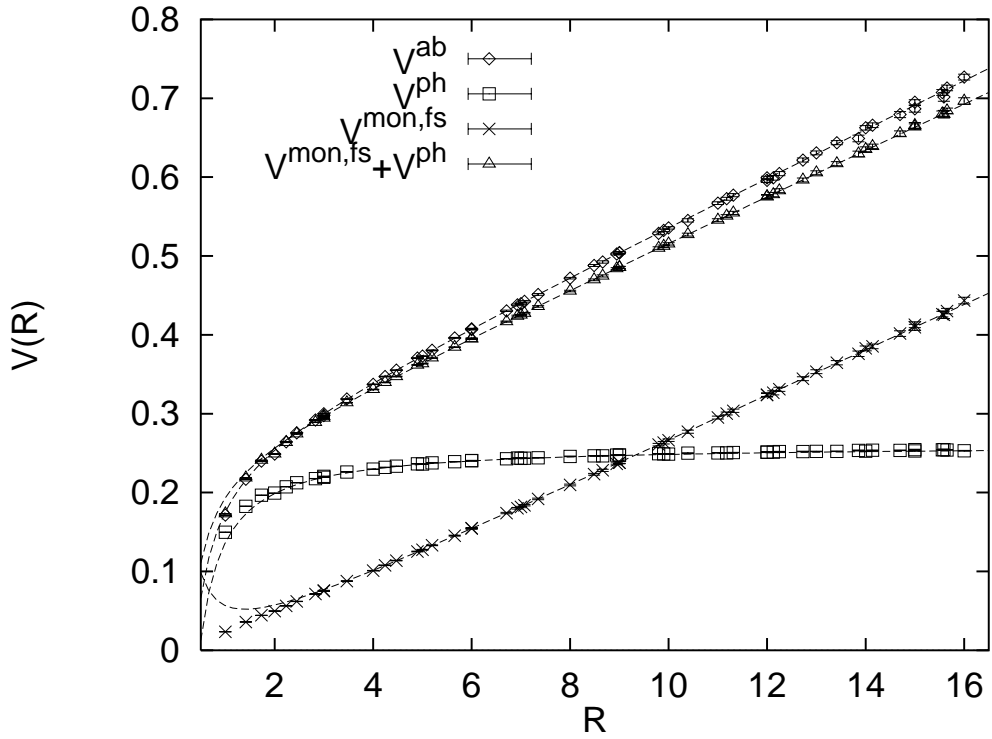}
\end{center}
\caption{The abelian potential (diamonds) in comparison with the photon
contribution (squares), the monopole (plus finite volume) contribution
(crosses) and the sum of these two parts (triangles). Notice, that no
(self energy) constants have been subtracted from any of the data sets.}
\label{fig:pot_decomp}
\end{figure}

The resulting potentials as well as $V^{ab}$ are displayed in
Fig.~\ref{fig:pot_decomp} and the parameter values
as obtained from fits according to Eq.~(\ref{potffit}) are quoted in
Table~\ref{t4}.
The corresponding values for the abelian potential are included as
well, where we have omitted the bias from gauge fixing ambiguities, to
allow for a direct comparison.

One can see from Fig.~\ref{fig:pot_decomp} that the photon part
$V^{ph}$ does not contribute to the string tension.  Therefore, in MA
projection only the monopole part of the abelian gauge fields,
$\theta^{mon}_{n,\mu}$, gives rise to the flux tube.  A comparison of
$V^{ab}$ with $V^{mon,fv}+V^{ph}$ reveals a qualitative
agreement, implying the approximate validity of the factorization
ansatz. The string tension of the monopole contribution amounts to
$(95\pm 1)\%$ of the full abelian string tension.  The approximate
decomposition of the abelian static potential into monopole and photon
parts gives evidence for the interaction term between monopoles and
photons within the corresponding effective action to be weak.

\begin{figure}[p]
\begin{center}
\leavevmode
\epsfxsize=14truecm\epsfbox{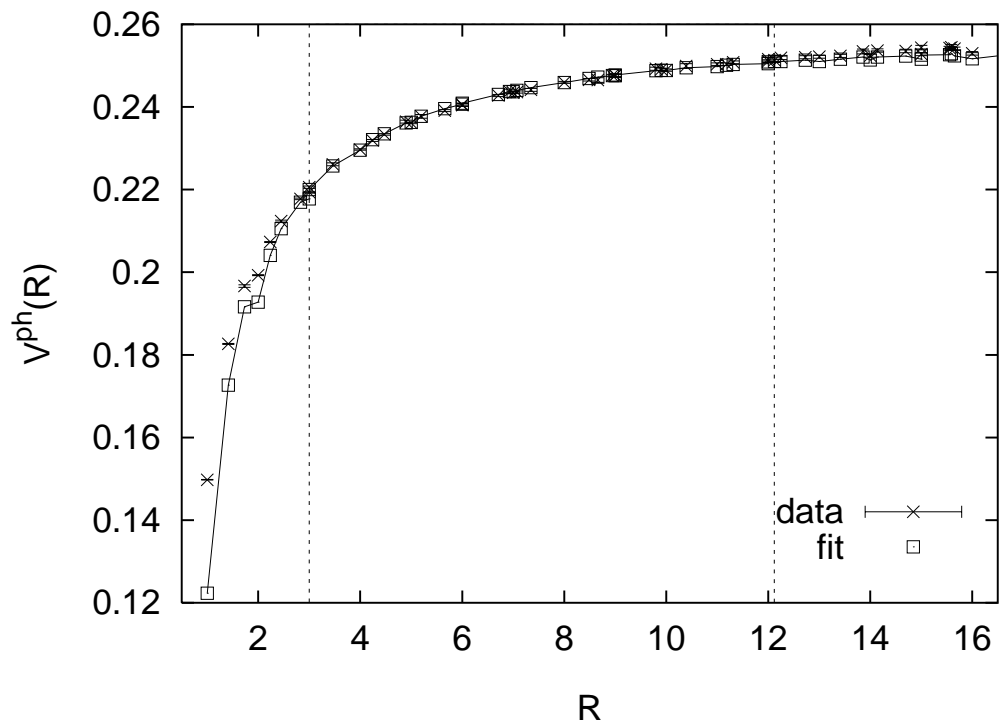}
\end{center}
\caption{The photon contribution to the abelian potential. The dashed
vertical lines indicate the fit range (solid curve and squares).}
\label{fig:photon}
\end{figure}

We have also attempted to
fit the photon part of the potential to the ansatz,
\eq
V^{ph}({\mathbf R})=V_{\mbox{\scriptsize self}}-fG_{L=32}({\mathbf R})
\quad,\quad G_L({\mathbf R})=\frac{4\pi}{V}\sum_{{\mathbf k}\neq{\mathbf 0}}
e^{i{\mathbf k}{\mathbf R}}D({\mathbf k},0)\quad
\en
for various fit ranges. $G_L({\mathbf R})$ is the Coulomb potential
on the lattice which
approaches $1/R$ in the infinite volume limit
($L\rightarrow\infty$) for large (lattice) $R$.
Lattice artefacts turned out to be well
parameterized by this functional form.
As a result we quote
$V_{\mbox{\scriptsize self}}=0.2513(3)$
and
$f=0.130(30)$, obtained on the fit range $3\leq R\leq 12.12$.
Data and fit curve are visualized in Fig.~\ref{fig:photon}.
The fit range turned out to be
correlated with the parameter value $f$ in so far as $f$ tended to be
larger if large $R$ values had been included and smaller for small $R$
values. This is also evident from the figure and might be interpreted
as a relict of asymptotic freedom within the abelian projected gauge
theory.
{}From tree level perturbation theory, one might expect
$f=V_{\mbox{\scriptsize self}}/G_{L=32}({\mathbf 0})\approx 0.0814$,
which is smaller than the fitted $f$ quoted above, in accordance with
a running coupling interpretation.

The monopole
contribution to
the abelian potential $V^{ab}_{l=2}$ has been extracted
{}from extended monopoles of size $l=2$ as well, by use of the method,
introduced in Section~2 (Eq.~(\ref{wlmonl})).
By computing the potential from the corresponding Wilson loops (which has
been done in Ref.~\cite{suz_monwl} the first time), we found the
approximants to decrease monotonously in $T$.
The value of the monopole string tension turned out to be slightly
smaller
than the one extracted from elementary monopoles
($K^{mon}_{l=2}=0.0271(3)$ instead of $K^{mon}=0.0290(3)$, c.f.\
Table~\ref{t4}),
which might indicate
that a small portion of the string tension
is due to monopole structures of smaller extent.

The new method of computing the monopole contribution to the potential
by combining Eqs.~(\ref{factor3}) and (\ref{factor2})
with the smearing method has the advantage of large ground state
overlaps. Also, the potential can easily be computed for off-axis
points while the blocking method even reduces the number of on-axis
points at which the potential can be measured by a factor two.
Notice, that $V^{mon,fs}$ obtained by use of the smearing method
is identical to $V^{mon}+V^{fs}$, as computed in the traditional way
{}from elementary monopole currents, while monopole structures on
the scale of a lattice spacing are
neglected in the blocking method.

\subsection{Charge two case}

\begin{figure}[p]
\begin{center}
\leavevmode
\epsfxsize=14truecm\epsfbox{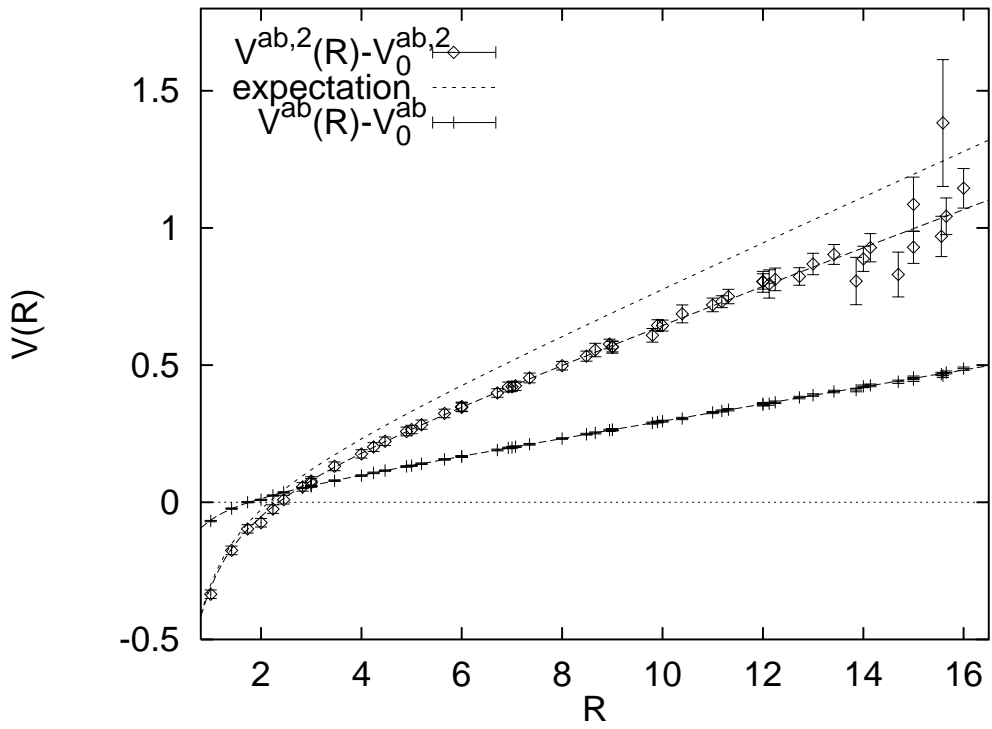}
\end{center}
\caption{The static abelian charge two potential $V^{ab,2}(R)$ in comparison
 to $V^{ab}(R)$. In addition,
the expectation
$V^{ab,2}(R)-V^{ab,2}_0\approx 8/3 K^{ab}R-4e^{ab}/R$ is included
(upmost dashed line).}
\label{fig:pot_ch2}
\end{figure}

In order to extract the charge two potential $V^{ab,2}(R)$
(Section~\ref{charge2}),
smeared charge two Wilson loops, $W^{ab,2}(R,T)$
have been
evaluated. The data for the fundamental and charge two abelian
potentials are displayed in Fig.~\ref{fig:pot_ch2} (with the
fitted self energies being subtracted).

Flux tube models lead to the expectation $K^{adj}/K=8/3$ for the ratio
of the adjoint over the the fundamental string tension\footnote{Of
course, the adjoint string tension is only an effective quantity since, at
large distance, string breaking is expected to set, caused by screening
from the creation of glueball pairs.}. This value
was qualitatively
supported by numerical data~\cite{ambjorn,michael}. However, recent
results~\cite{michael2,trott}
indicate that at intermediate to large distances the above ratio
tends to be somewhat smaller than $8/3$.
We suggest the slope of the charge two potential, $K^{ab,2}$
to constitute the abelian projection counterpart to the adjoint string
tension and thus expect
the ratio
$K^{ab,2}/K^{ab}$ to agree with results on $K^{adj}/K$.
However, the lack of high precision data on the latter ratio
prevents us from a quantitative test of this assumption.
{}From our data, we find the value $K^{ab,2}/K^{ab}=2.23(5)$
significantly smaller than 8/3.

For small $R$, perturbation theory
yields the same $8/3$ ratio between the adjoint and the
fundamental potential, which differs from the expected ratio between abelian
charge two and charge one potentials, where one naively
would expect a factor $2^2=4$.
We indeed obtain $V_0^{ab,2}/V_0^{ab}= 3.96(7)$ and $e^{ab,2}/e^{ab}=4.0(5)$
for the fit parameters that are sensitive to short range physics.
We have included the curve $8/3K^{ab}-4e^{ab}/R$ into
Fig.~\ref{fig:pot_ch2}, which corresponds to
$SU(2)$ Casimir scaling for the linear part of the potential and
to the perturbative $U(1)$ expectation for its Coulomb part.

\subsection{Monopole density}

Monopole densities for elementary
as well as for extended monopole currents of size $l=2$ have been
evaluated.
The density of monopoles of size $l$ in physical units is
defined as\footnote{The factor $l^3$ appears
in the denominator of Eq.~(\ref{monden}), due to
averaging over the $l^3$ possible
blocked sublattices.}
\eq
\rho^{(l),ph}_{\mbox{\scriptsize mon}} =
\frac{1}{4V \cdot (la)^3} \sum_{n,\mu}\langle|k^{(l)}_{n,\mu}|\rangle\quad.
\label{monden}
\en
\noi
We obtain the following results for the monopole densities,
converted into units of the string tension as measured in
the present investigation:
\eq
\frac{\rho^{ph}_{\mbox{\scriptsize mon}}}{K^{3/2}} = 1.962(4)
\quad,\quad
\frac{\rho^{(2),ph}_{\mbox{\scriptsize mon}}}{K^{3/2}} = 1.269(2)
\quad.
\label{mondenres}
\en
The value for elementary monopole currents has been corrected by its
systematic bias from incomplete gauge fixing.

In order to relate our results to those obtained in
previous publications,
we compare our value on the density of elementary monopoles
with the value $\rho^{ph}_{\mbox{\scriptsize mon}}/K^{3/2} = 2.11(2)$
from Ref.~\cite{big} for a $16^4$ lattice
at $\beta=2.5$, which
is close to our coupling constant, $\beta=2.5115$. Note, that we have
rescaled the result of Ref.~\cite{big} into units of the string tension
as obtained in Ref.~\cite{smearing}.
The result of Ref.~\cite{big} turns out to be
consistent within errors with results from other
authors~\cite{suzener,stwen_su2,dgdens} while our value is
significantly smaller. Since finite size effects are negligible for
the lattice extents under consideration,
the difference seems to be due to our improved
OSA gauge fixing algorithm.
A similar discrepancy is observed for
extended monopoles. In this case our value, Eq.~(\ref{mondenres}),
should be compared with
$\rho^{(2),ph}_{\mbox{\scriptsize mon}}/K^{3/2} = 1.32 $,
as obtained on a $24^4$ lattice at $\beta=2.5$ by means of
the standard OR gauge fixing procedure~\cite{shiba}.

\section{Summary and conclusions}

Let us summarize our main results and conclude:

\begin{itemize}

\item The present study is based on a self contained and self
consistent analysis on the largest lattice volume that has been
studied so far for this kind of simulations with comparatively high
statistics. A systematic error analysis has been
carried out for the first time in this context.
\item
To obtain reliable results in the abelian projected theory
with MA gauge condition, one
has to investigate and control the uncertainty that is inevitably
introduced by the incomplete gauge fixing of numerical practice.
The OSA algorithm has been shown to be a powerful tool for gauge fixing.
A method for estimation of residual uncertainties is proposed.

\item We have found $K^{ab}$ to be $(8\pm 4) \%$ smaller than the
non-abelian string tension at $\beta=2.5115$.

\item
Our investigation of the decomposition of the static abelian potential
into monopole and photon parts confirms earlier observations at a
higher confidence level. By applying a new method for extracting the
monopole contribution to the potential, we have been able to extract
the corresponding potentials and fit parameters reliably.  The
factorization has been found to work qualitatively, the monopole
contribution accounting for the string tension within a margin of $5
\%$.

\item
We have calculated the abelian projection approximation for the adjoint
string tension. Our result for ratio
$K^{ab,2}/K^{ab}$ is in qualitative agreement with numerical data on
its non-abelian counterpart.

\end{itemize}

We believe that further computations at different $\beta$ values,
provided all sources of errors are kept under control
as in the present
paper, will answer the question whether the abelian
projected
theory exactly reproduces the large distance behavior of the full
theory in the continuum limit.

\section{Acknowledgements}

We thank the Deutsche Forschungsgemeinschaft for supporting the
Wuppertal CM-5 project (grants Schi 257/1-4 and Schi 257/3-2)
and the HLRZ for computing
time on the CM-5 at GMD. GB, KS and MMP appreciate support
by European Union contracts
SC1*-CT91-0642 and CHRX-CT92-0051. During completion
of this work GB received funding by EU contract ERB-CHBG-CT94-0665.
This work was supported in part by grant No.\ NJP000, financed by
the International
Science Foundation, by grant No.\ NJP300, financed by
the International Science Foundation and by the Government of the
Russian Federation.
VB obtained partial support from DFG grant 436-RUS-113
and from Grant No.\ 93-02-03609, financed by
the Russian Foundation for Fundamental Sciences.

\eject
\noindent {\large\bf Note added in proof}
\vskip .5cm

\noindent
After completion of the present paper, a preprint by
G.~Poulis~\cite{gpoulis} has been received, in which
the author has demonstrated that one should expect
$K^{ab,2}$ and $K^{adj}$ to coincide to the same extent as
$K^{ab}$ and $K$ agree by use of reasonable
approximations.

\begin{table}[p]
\caption{Comparison of gauge fixing algorithms.}\label{t1}
\setlength{\tabcolsep}{0.55pc}
\begin{tabular}{ccccc}  \hline
       &  RO        & OSA1     & OSA2     &  OSA3      \\
\hline
$\langle \overline{F} \rangle$&0.7370(2)      &0.7383(2)   & 0.7387(2)  &   0.7390(2)               \\
$\delta_{sd}$& $18 \cdot 10^{-5}$&$13 \cdot 10^{-5}$&$10 \cdot 10^{-5}$&
$8 \cdot 10^{-5} $    \\
$\rho_{\mbox{\scriptsize mon}}$
& 0.0218(2) & 0.0209(3)& 0.0207(3) & 0.0204(3) \\
$W_{11}   $ & 0.7702(5) & 0.7716(5)& 0.7720(5) & 0.7723(5) \\
$W_{44}   $ & 0.085(1)  & 0.090(1) & 0.091(1)  & 0.092(2)  \\
$K^{ab}$&0.063(3)&0.057(3)&0.055(3)&0.054(3)\\
\hline
\end{tabular}
\end{table}

\begin{table}[p] \caption{Dependence of results on the number of
gauge copies, $m$, at
$\beta=2.5115$, $V=32^4$, $N_c=30$.}\label{t2}
\setlength{\tabcolsep}{0.55pc}
\begin{tabular}{ccccc}  \hline
&$m=1$&$m=10$&$m=20$&$m=\infty$\\
\hline
$\langle F \rangle$&
0.752641(34)&0.752789(35)&0.752822(36)&0.752845(37)\\
$S$&
0.202508(60)&0.202299(64)&0.202267(64)&0.202261(66)\\
$\rho_{\mbox{\scriptsize mon}}$&
0.011639(26)&0.011528(30)&0.011515(32)&0.011513(33)\\
$K^{ab}$&0.0325(11)&0.0311(13)&0.0306(16)&0.0303(17)\\\hline
\end{tabular}
\end{table}

\begin{table}[p] \caption{Fit parameters for the static potentials in
abelian projected $SU(2)$ (corrected by the estimated bias due to
incomplete gauge fixing) and the potential of full $SU(2)$. Errors are
systematic and statistical.}\label{t3}
\setlength{\tabcolsep}{0.55pc}
\begin{tabular}{cccc}  \hline
       &  AP $SU(2)$  & $SU(2)$       \\
\hline
$e    $ &$ 0.095(11) $ &$ 0.252(24) $              \\
$V_0   $ &$ 0.240(4)  $ &$0.545(10)   $   \\
$K$ &$ 0.0297(8)$ &$ 0.0325(12)$  \\
\hline
\end{tabular}
\end{table}

\begin{table}[p] \caption{Test of factorization of the abelian
potential into monopole and photon parts. Biases from incomplete gauge
fixing have been omitted throughout the table.}\label{t4}
\setlength{\tabcolsep}{0.55pc}
\begin{tabular}{cccccc}  \hline
     & $V^{ab}$&$V^{mon,fv}+V^{ph}$&$V^{mon,fv}$&$V^{mon}_{l=2}$&$V^{ph}$\\
\hline
$e$  &0.095(6) &0.068(5) &-0.056(5) &0.019(4) &0.124(2)  \\
$V_0$&0.240(3) &0.232(2) &-0.029(2) &0.010(2) &0.261(1)  \\
$K$  &0.0305(3)&0.0291(3)&0.0290(3) &0.0271(3)&0.00007(4)\\
\hline
\end{tabular}
\end{table}

\begin{table}[p] \caption{Static potential parameters for
the abelian projections of the fundamental $(q=1)$
and adjoint $(q=2)$ representations of $SU(2)$.}\label{t5}
\setlength{\tabcolsep}{0.55pc}
\begin{tabular}{cccc}  \hline
       &  AP $SU(2)~ (q=1)$  & AP $SU(2)~ (q=2)$&$(q=2)/(q=1)$\\
\hline
$e    $ &$ 0.095(6)$ &0.376(36)&3.98(48)    \\
$V_0   $ &$ 0.240(2)$ &0.950(13)&3.96(7)     \\
$K$ &$0.0305(3) $ &0.0682(11)&2.23(5)   \\
\hline
\end{tabular}
\end{table}

\end{document}